\documentclass{aa}  
\usepackage{graphicx}
\usepackage{amsmath}
\usepackage{amssymb}
\usepackage{txfonts}
\newcommand{\ros}{{\it ROSAT}}
\newcommand{\chan}{{\it Chandra}}
\newcommand{\xmm}{{\it XMM-Newton}}
\newcommand{\eros}{{eROSITA}} 
\newcommand{\srgl}{{\it Spectrum Roentgen Gamma}}
\newcommand{\srg}{{\it SRG}}
\newcommand{\nicer}{{\it NICER}}
\newcommand{\nicerl}{{Neutron star Interior Composition Explorer}}
\newcommand{\eso}{{\it ESO-VLT}}
\newcommand{\forst}{{FORS2}}

\def \magoe{\object{RX~J1856.5--3754}}
\def \magzs{\object{RX~J0720.4--3125}}
\def \magos{\object{RX~J1605.3+3249}}
\def \magot{\object{RX~J1308.6+2127}}
\def \magto{\object{RX~J2143.0+0654}}

\def \magzf{\object{RX~J0420.0--5022}}
\def \jotos{\object{eRASSU~J131716.9--402647}}
\def \otos{\object{J1317}}
\def \zsts{\object{PSR~J0726--2612}}

\def \fluxcgs{erg~s$^{-1}$~cm$^{-2}$}
\newcommand{\nh}{N_{\rm H}}

\begin{document} 
\title{A phase-coherent timing solution for the X-ray dim isolated neutron star \jotos
\thanks{Based on observations obtained with \xmm, an ESA science mission with instruments and contributions directly funded by ESA Member States and NASA (observations 0921280101)}}
\author{J.~Kurpas\inst{1}
\and A.~M.~Pires\inst{2,1}
\and A.~D.~Schwope\inst{1}
\and F.~Haberl\inst{3}
\and S.~Sheth\inst{1,4}
}
\offprints{J. Kurpas}
\institute{Leibniz-Institut f\"ur Astrophysik Potsdam (AIP), An der Sternwarte 16, 14482 Potsdam, Germany
\email{jkurpas@aip.de} 
\and
Center for Lunar and Planetary Sciences, Institute of Geochemistry, Chinese Academy of Sciences, 99 West Lincheng Rd.,\\550051 Guiyang, China
\and
Max-Planck-Institut f\"ur extraterrestrische Physik, Gie{\ss}enbachstra\ss e 1, 85748 Garching, Germany
\and
Potsdam University, Institute for Physics and Astronomy, Karl-Liebknecht-Stra\ss e 24/25, 14476 Potsdam, Germany
}
\date{Received ...; accepted ...}
\titlerunning{A coherent timing solution for \otos}
\authorrunning{Kurpas~J., et al.}
\abstract
{
Based on its predominantly thermal X-ray emission and long spin period, the isolated neutron star (INS) \jotos\ is one of the most promising candidates for membership in the still small class of X-ray dim isolated neutron stars (XDINSs). Confirmation of this classification, however, requires a more detailed characterisation of the source's timing and spectral properties. In this work, we present new \nicer\ observations which, together with previous X-ray follow-up, allow us to constrain the timing properties and long-term evolution of \jotos. We obtain a coherent timing solution with a spin period of $P\sim12.8$~s and a period derivative of $\dot{P}\sim9\times 10^{-14}$~s~s$^{-1}$, which best-describes the spin evolution of the source. These parameters imply a dipolar magnetic field strength of $3\times10^{13}$~G and a spin-down luminosity of order $10^{30}$~erg~s$^{-1}$. Spectral modelling reveals no significant change in the spectral state over the 15 months of observational monitoring and indicates a thermal luminosity that likely exceeds the rotational energy loss. This suggests a thermal evolution that has been significantly influenced by past reheating. The energy dependence of the double-humped pulse profile closely resembles that observed in the XDINS \magot, with the pulsed fraction increasing towards higher energies. Taken together, these results unambiguously confirm the XDINS nature of \jotos, making it the first newly confirmed XDINS in more than two decades.
}
\maketitle
\section{Introduction\label{sec_intro}}

\begin{table*}
\small
\caption{Summary of all X-ray observations of \otos.\label{tab_obs}}
\centering
\scalebox{1.}{
\begin{tabular}{lclcrrcc}
\hline\hline\noalign{\smallskip}
Telescope & Instrument & OBSID & MJD\tablefootmark{(a)} & GTI\tablefootmark{(b)} & \multicolumn{1}{c}{Counts} & Bkg. Percentage\tablefootmark{(c)} & Energy Band\tablefootmark{(d)} \\
& & & [days] & [ks] & & [\%] & [keV]\\
\hline
\nicer & XTI       & 65720201(01--04) & 60034.018840 & 41.4 & 26419 & $36.8\pm0.4$ & 0.55--1.20\\
\xmm   & EPIC MOS1 & 0921280101      & 60133.786367 & 37.8 &  1819 & $0.5\pm2.4$ & 0.20--2.00 \\
\xmm   & EPIC MOS2 & 0921280101      & 60133.786486 & 38.1 &  1960 & $0.6\pm2.3$ & 0.20--2.00 \\
\xmm   & EPIC pn   & 0921280101      & 60133.798929 & 31.7 &  9254 & $0.9\pm1.1$ & 0.20--2.00 \\
\nicer & XTI       & 76170101(01--06) & 60395.232585 & 34.3 & 18992 & $40.6\pm0.5$ & 0.55--1.00\\
\nicer & XTI       & 76170102(01--02) & 60403.610537 &  9.3 &  3301 & $38.3\pm0.8$ & 0.65--1.20\\
\nicer & XTI       & 76170103(01--03) & 60431.225607 & 12.4 &  6305 & $49.0\pm0.8$ & 0.60--1.35\\
\nicer & XTI       & 76170104(01--03) & 60485.033690 & 15.1 &  7147 & $41.1\pm0.7$ & 0.60--1.15\\
\hline
\end{tabular}}
\tablefoot{
\tablefoottext{a}{Modified Julian date at mid-point of the observation.}
\tablefoottext{b}{Remaining 'good' observing time after applying the data reduction procedures and screening for periods of high background activity (see text for details).}
\tablefoottext{c}{Percentage of detected events attributed to the background. For \nicer, this is estimated from the best-fit SCORPEON model. For \xmm, the percentage is computed from the counts in the defined background region, rescaled to match the source region.}
\tablefoottext{d}{Energy band used for the timing analysis. Counts and background percentage columns refer to this band.}

}
\end{table*}

X-ray dim isolated neutron stars (XDINSs) are characterised by predominantly thermal X-ray spectra that are largely uncontaminated by magnetospheric emission, providing a direct view of their cooling surfaces \citep{2009ASSL..357..141T, 2022MNRAS.516.4932D}. Originally discovered with \ros\ observations \citep{2007Ap&SS.308..181H}, the seven confirmed members of this class (also known as the 'Magnificent Seven') exhibit long spin periods ($\sim3-12$~s), strong magnetic fields ($B\sim 10^{13}-10^{14}$~G), and they are relatively nearby, with distances $<1$~kpc  \citep[see][and references therein]{2024ApJ...969...53B, 2007Ap&SS.308..171P, 2010MNRAS.402.2369T}.

Despite being far less numerous than rotation-powered pulsars \citep[RPPs;][]{2009ASSL..357...91B}, XDINSs have a local spatial density comparable to that of the RPP population and they may therefore represent a substantial fraction of the overall isolated neutron star (INS) population \citep{2008MNRAS.391.2009K}. Their thermal X-ray emission also makes XDINSs particularly valuable laboratories for testing magneto-thermal evolutionary models, which suggest an evolutionary link with the magnetar population \citep{2013MNRAS.434..123V, 2017ARA&A..55..261K}. In addition, they provide important constraints on neutron star cooling and on the physical state and composition of their surface layers \citep[e.g.][]{2020MNRAS.496.5052P,2007Ap&SS.308..191V,2007MNRAS.375..821H}. Furthermore, XDINSs offer a unique opportunity to probe neutron star structure. Measurements of absorption features and radius estimates enable constraints on the stellar compactness and, in turn, on the equation of state of cold nuclear matter \citep[e.g.][]{2011A&A...534A..74H,2016ARA&A..54..401O, 2024APh...15802935A}. Consequently, constraining the intrinsic properties, population characteristics, and Galactic-scale evolution of XDINSs is of considerable interest, but it remains limited by the small number of known sources.

One of the most promising XDINS candidates to be discovered in recent years is the soft, thermally emitting INS \jotos\ (henceforth dubbed \otos). The source was originally selected from the \srgl\ (\srg)/\eros\ All-Sky Survey \citep{2021A&A...647A...1P,2023A&A...674A.155K,2024A&A...682A..34M}. Subsequent X-ray follow-up with \xmm\ \citep{2001A&A...365L...1J} and the \nicerl\ \citep[\nicer;][]{2016SPIE.9905E..1HG}, together with deep optical imaging obtained with \forst\ at the \eso\ \citep{1998Msngr..94....1A}, established its INS nature through the absence of detectable optical counterparts and the discovery of a spin period of $\sim12.8$~s \citep{2024A&A...683A.164K}.

X-ray spectral modelling showed that the data are best described by a single blackbody component with $kT\sim90$~eV and two absorption features at $\sim250$~eV and $\sim590$~eV. The blackbody-like continuum and the inferred magnetic field strength of $\sim5\times10^{13}$~G, assuming the observational feature at $\sim250$~eV arises from proton cyclotron resonances, are consistent with properties observed in known XDINSs and high-magnetic field pulsars \citep{2011AIPC.1379...60N,2019A&A...627A..69R}.

Despite these similarities, the exact classification of \otos\ within the Galactic INS population remains uncertain, motivating further observations. In particular, a measurement of the spin-down rate and the construction of a phase-coherent timing solution are required to independently constrain the dipolar magnetic field strength and to firmly place the source among the known INS classes. To this end, four additional \nicer\ observations were obtained in 2024 with the goal of better constraining the timing properties and long-term evolution of the source. In this paper, we report on the results of this observational campaign.

The paper is structured as follows. In Sect.~\ref{sec_obs} we describe the data reduction of the new and archival \nicer\ and \xmm\ observations of \otos. In Sect.~\ref{sec_analysis}, we present the results of the timing analysis and spectral modelling. Finally, in Sect.~\ref{sec_disc} and \ref{sec_concl}, we discuss the results and summarise our conclusions.

\section{Observations and data reduction\label{sec_obs}}

\subsection{\nicer\label{sec_obsnic}}

The source \otos\ was observed with the \nicer\ instrument over five epochs between March 2023 and June 2024, with each epoch consisting of multiple individual pointings (Table~\ref{tab_obs}). Data reduction was performed using the NICERDAS software (version: 2025-06-11\_V014) that is distributed within the HEASOFT package (version: 12Mar2025\_V6.35), together with the most recent \nicer\ CALDB files (version: XTI20240206).

The reduction was carried out by running the \texttt{nicerl2} task on all individual OBSIDs, adopting conservative screening criteria. Specifically, we set the \texttt{saa\_filt} parameter to \texttt{yes} and \texttt{nicersaafilt} to \texttt{no}, used an \texttt{under\_only\_range} of 0--50,  an \texttt{over\_only\_range} of 0--5, applied a \texttt{noise\_ring\_under} threshold of 80, a \texttt{low\_mem} value of 250, an \texttt{elv\_lim} of 30, \texttt{cor\_range} of \texttt{$1.5-\star$}, and the \texttt{NICERV6} filter column. Daytime and nighttime data were processed separately, although the majority of observations were obtained during orbital night. Only for epochs 76170101(01--06) and 76170103(01--03) did orbit-day data pass the initial screening, yielding a limited number of events (626 and 796 counts between 0.3--2~keV, respectively). Consequently, the bulk of the \nicer\ data is unaffected by the optical light-leak contamination\footnote{More information on the light leak is available at: \url{https://heasarc.gsfc.nasa.gov/docs/nicer/analysis_threads/light-leak-overview/}}.

Observations belonging to the same epoch were merged using the \texttt{niobsmerge} task. We then created a 30~s binned light curve in the 0.3--12~keV band and applied a $\sigma$-clipping algorithm to the light curve to filter periods of enhanced background activity, rejecting time intervals where the count rate deviated by more than $3\sigma$ from the epoch-averaged mean. In addition, time intervals corresponding to solar $k_P$ index values greater than five were manually excluded. The event times were corrected to the solar-system barycentre using the \texttt{barycorr} task, adopting the source position (RA: 13:17:17.03, DEC: -40:26:46.90) derived from the \xmm\ observation reported by \citet[][]{2024A&A...683A.164K} and the DE405 ephemeris. Next, we assigned phase values to individual events using the \texttt{photonphase} Python script from the \texttt{NICERSOFT} package\footnote{Available at: https://github.com/paulray/NICERsoft}, adopting the spin period reported by \citet{2024A&A...683A.164K}. Optimal energy bands for pulsation detection were subsequently defined using the \texttt{nioptcuts} script and are listed in Table~\ref{tab_obs}. Event lists extracted in these energy bands were used for the timing analysis.

For the spectral analysis, source and SCORPEON background spectra were extracted for each epoch using the \texttt{nicerl3-spect} task. The resulting spectra were grouped with \texttt{ftgrouppha} to ensure a minimum of 25 counts per spectral bin.

\subsection{\xmm\label{sec_obsxmm}}

\xmm\ observed \otos\ in July 2023 (Table~\ref{tab_obs}). Equipped with \texttt{THIN} filters, EPIC pn \citep{2001A&A...365L..18S} was operated in full frame mode and EPIC MOS1 and MOS2 \citep{2001A&A...365L..27T} in small window mode. Data reduction generally followed the procedures described in \citet{2024A&A...683A.164K,2026A&A...705A.148K}. The Science Analysis System (SAS; version: 22.1.0) was used, selecting only single and double pattern events for EPIC pn (PATTERN$\leq 4$) and all valid patterns for EPIC MOS (PATTERN$\leq 12$) that are uncorrupted (FLAG$=0$).

Periods of high background were identified and removed using a $\sigma$-clipping algorithm to a 50~s binned background light curve, extracted from the full chip in the 0.3--12~keV band after masking bright sources based on a prior \texttt{edetect\_stack} run. Background regions were defined on the same chip and RAWY coordinate as the source for EPIC pn and on a neighbouring chip for the MOS detectors. Source regions were optimised for signal-to-noise using the \texttt{eregionanalyse} task.

For the timing study, event lists were extracted from the source region in the 0.2--2~keV range and corrected to the solar-system barycentre using the \texttt{barycen} task, the DE405 ephemeris, and the source position reported by \citet{2024A&A...683A.164K}. Source spectra were also extracted following standard SAS guidelines and grouped to a minimum of 25 counts per spectral bin, with a maximum oversampling factor of three.

\section{Results\label{sec_analysis}}

\subsection{Timing analysis \label{sec_timing}}

\begin{table}
\caption{Timing parameters.\label{tab_tim}}
\centering
\scalebox{1.}{
\begin{tabular}{lr}
\hline\hline\noalign{\smallskip}
\multicolumn{2}{l}{Timing solution}\\
\hline
Reference time [MJD] & 60392.848396\\
Frequency [Hz] & $0.0783875062(14)$\\
Frequency derivative [s$^{-2}$] & $-5.4^{+1.0}_{-1.1} \times 10^{-16}$\\
Period [s] & $12.75713501(23)$\\
Period derivative [ss$^{-1}$] & $8.8^{+1.7}_{-1.6} \times 10^{-14}$\\
Phase offset\tablefootmark{(a)} & $-4^{+5}_{-5}\times 10^{-3}$\\
\hline
\multicolumn{2}{l}{Related parameters}\\
\hline
Dipolar magnetic field strength\tablefootmark{(b)} [G] & $3.4^{+0.4}_{-0.4} \times 10^{13}$\\
Characteristic age\tablefootmark{(c)} [Myr] & $2.3^{+0.5}_{-0.4}$\\
Spin-down luminosity\tablefootmark{(d)} [erg s$^{-1}$] & $2.7^{+0.6}_{-0.5}\times 10^{30}$\\
\hline
\end{tabular}}
\tablefoot{
\tablefoottext{a}{The phase offset refers to the $\phi_0$ variable in Eq.~\ref{eq_time_sol}.}
\tablefoottext{b}{Computed from $B_\mathrm{dip} = 3.2\times 10^{19}\sqrt{P\dot{P}}$.}
\tablefoottext{c}{The characteristic age is derived, using $\tau = P\times(2\dot{P})^{-1}$.}
\tablefoottext{d}{We assumed a canonical INS with $M=1.4$~M$_\odot$ and $R=12$~km, such that the luminosity follows $\dot{E} = 6.34\times 10^{46}(\dot{P}P^{-3})$~g~cm$^2$.}
}
\end{table}

We began the timing analysis of \otos\ by defining the reference time for phase zero. To this end, we applied the Bayesian folding method of \citet{1996ApJ...473.1059G} to the second \nicer\ epoch (76170101(01--06), see Table~\ref{tab_obs}) to measure the source's pulsation. We detected the INS modulation at a period of $12.757133(14)$~s with very high significance, corresponding to nearly 100\% likelihood that the pulsation is present in the data. Using this period, we generated a phase-folded light curve and set the time of phase zero to coincide with the deepest minimum of the double-humped pulse profile, which represents the most characteristic feature of the modulation. This defines $t_0 = 60392.848396$, close to the start of the second \nicer\ epoch.

With the reference time defined, we searched for a phase-coherent timing solution that best-fits all \otos\ observations, using the same algorithm applied in \citet{2025A&A...694A.160K}. The goal is to determine the timing parameters that minimise the residuals between observed and predicted cycle counting, defined as
\begin{equation}
\label{eq_toares}
R = \sum_{i=1}^{N} \left(\frac{\phi_\mathrm{obs,i} - \phi(t_i,t_0,\nu,\dot{\nu},\phi_0)}{\sigma(\phi_\mathrm{obs,i})}\right)^2,
\end{equation}
where $N$ is the number of individual times-of-arrival (TOA) considered.
The cycle counts are modelled using a linear or quadratic timing solution of the form
\begin{equation}
\label{eq_time_sol}
\phi(t,t_0,\nu,\dot{\nu},\phi_0) = \phi_0+\nu(t-t_0)+\frac{1}{2}\dot{\nu}(t-t_0)^2,
\end{equation}
where $\nu$ and $\dot{\nu}$ are the spin frequency and its derivative, $t_0$ is the reference time, and $\phi_0$ is a phase offset.

To increase the number of TOA measurements and more accurately track \otos's timing evolution, we subdivided the individual \nicer\ epochs into segments containing at least 4000 counts, while including the \xmm\ epochs as single segments. In total, this procedure defined 15 individual segments (see Appendix~\ref{sec_appa} for details).

For each segment and test timing solution, the phase-folded light curve was computed using 25 phase bins. The resulting pulse profile was then fitted with a sinusoidal function including both the first and second harmonics. We find a two harmonical sinusoidal function to be sufficiently accurate, as the fit quality barely improves, if higher orders are included in the pulse profile modelling (the fit statistic ranges from $\chi^2_\nu(\nu) = 0.61-1.58(20)$ and $\chi^2_\nu (\nu)= 0.57-1.60(18)$, if the pulse profiles of the individual segments are computed from the quadratic timing solution reported in Table~\ref{tab_tim} and they are fitted with two or three harmonics, respectively). From the fitted model, we determined the phase corresponding to the deepest minimum ($\phi_\mathrm{min}$) of the pulse profile, adopting a phase uncertainty of 0.02. Also here, the two harmonical fit is sufficiently accurate, as we find the resulting $\phi_\mathrm{min}$ values to agree generally well with those derived from pulse profile fits including also higher orders. By inverting Eq.~\ref{eq_time_sol} and using the trial timing parameters, we calculated the time and associated error of the pulse minimum closest to the mid-point of the segment. This time is then defined as the TOA for the segment.

We explored the likelihood landscape defined by Eq.~\ref{eq_toares} and determined the best-fitting timing solution using the reactive nested sampling algorithm implemented in the \texttt{UltraNest} package \citep{2021JOSS....6.3001B}. In this search, 800 sample points were evolved, uniform priors were assumed for the spin frequency and phase offset and a log-uniform prior was adopted for the spin-down frequency, to avoid introducing bias. The sampling terminated when the unexplored fraction of the evidence integral fell below 0.001\%.

\begin{figure}[t]
\includegraphics[width=\linewidth]{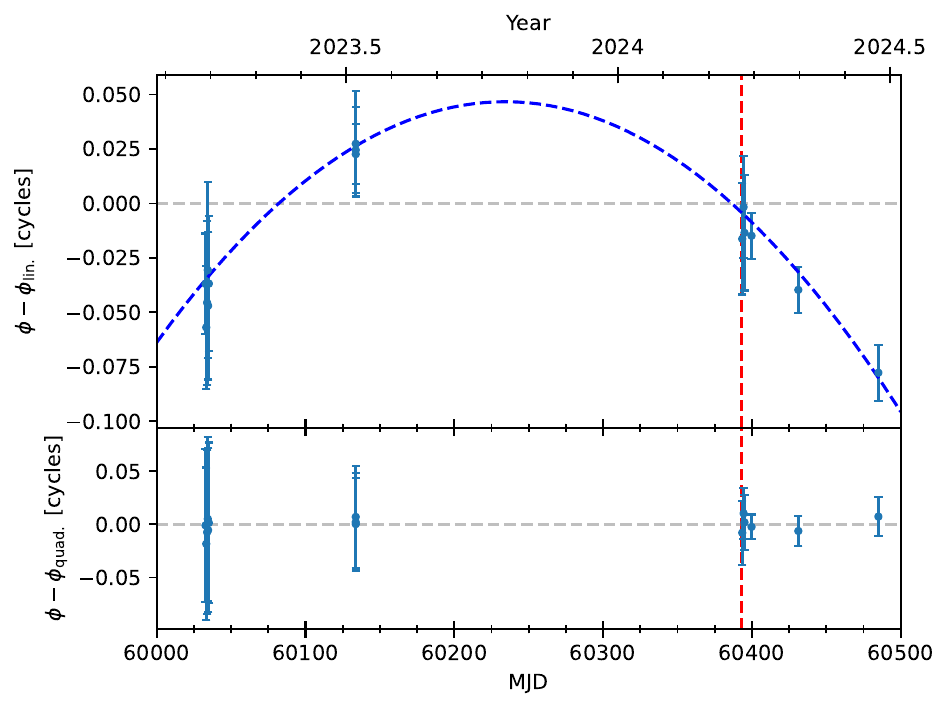}\\
\caption{Cycle-count comparison for the linear (top panel) and quadratic (bottom panel) timing solutions derived from TOA fitting. The blue dashed line in the top panel indicates the deviation between the linear and quadratic solutions, while the vertical red dashed line marks the reference time $t_0=60392.848396$.}
\label{fig_tsol_comp}
\end{figure}


Starting with a linear timing solution, we determined a spin period of $P=12.75713380(6)$~s and a phase offset of $\phi_0=-0.0110^{+0.004}_{-0.005}$. Throughout this paper, absolute values correspond to the point of lowest timing residuals, whereas errors are derived from the 15.9 and 84.1 percentile values of the posterior probability distributions for the respective parameters.

\begin{figure}[t]
\includegraphics[width=\linewidth]{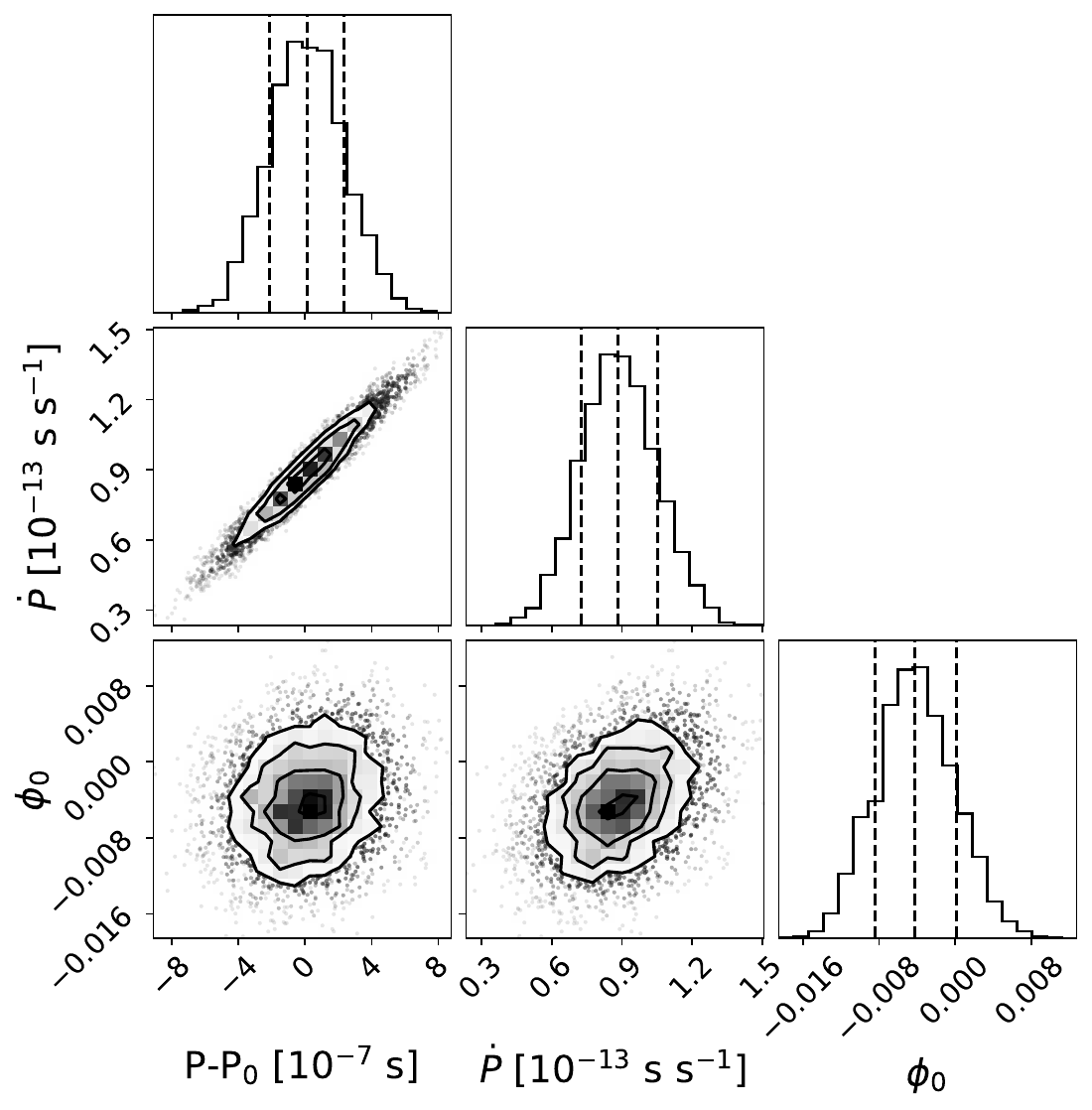}\\
\caption{Corner plot showing the posterior distributions of the parameters obtained from fitting a quadratic timing solution to the TOAs of the individual X-ray observations. For clarity, a reference period of $P_0 = 12.757135$~s has been subtracted from the posterior values.}
\label{fig_corn_plot}
\end{figure}


The predicted cycle counts from this linear solution, however, were found to deviate significantly from the observed values (Fig~\ref{fig_tsol_comp}), indicating that \otos's period evolves over the course of the observational campaign. To account for this, we next fitted a quadratic timing solution, which coherently connects all X-ray observations presented here, allows for accurate cycle counting, and successfully models the spin-evolution of the source (Fig~\ref{fig_tsol_comp}). From the posterior probability distributions of the quadratic solution (Fig.~\ref{fig_corn_plot}), we obtained the following timing parameters, $P=12.75713501(23)$~s, $\dot{P}=8.8^{+1.7}_{-1.6} \times 10^{-14}$~s~s$^{-1}$ and $\phi_0 = -4^{+5}_{-5}\times 10^{-3}$ (see also Table~\ref{tab_tim}). The resulting TOAs and cycle counts are reported in Appendix A. Given the excellent agreement of the quadratic solution with the observed TOAs, a first order correction from a linear timing solution is sufficient to model the observed spin evolution of \otos.

With the measurement of \otos's spin-down rate, it is now possible to estimate its dipolar magnetic field strength and characteristic age, under the assumption that magnetic dipole radiation dominates the spin evolution of the source \citep{1969ApJ...157.1395O}. Within this framework, we used the posterior probability distributions of the timing parameters to derive a dipolar magnetic field strength of $B= 3.4^{+0.4}_{-0.4} \times 10^{13}$~G and a characteristic age of $\tau=2.3^{+0.5}_{-0.4}$~Myr. Assuming a canonical neutron star with a mass of 1.4~M$_\odot$ and a radius of 12~km, we further obtained a spin-down luminosity of $\dot{E} = 2.7^{+0.6}_{-0.5}\times 10^{30}$~erg~s$^{-1}$.

\begin{figure}[t]
\includegraphics[width=\linewidth]{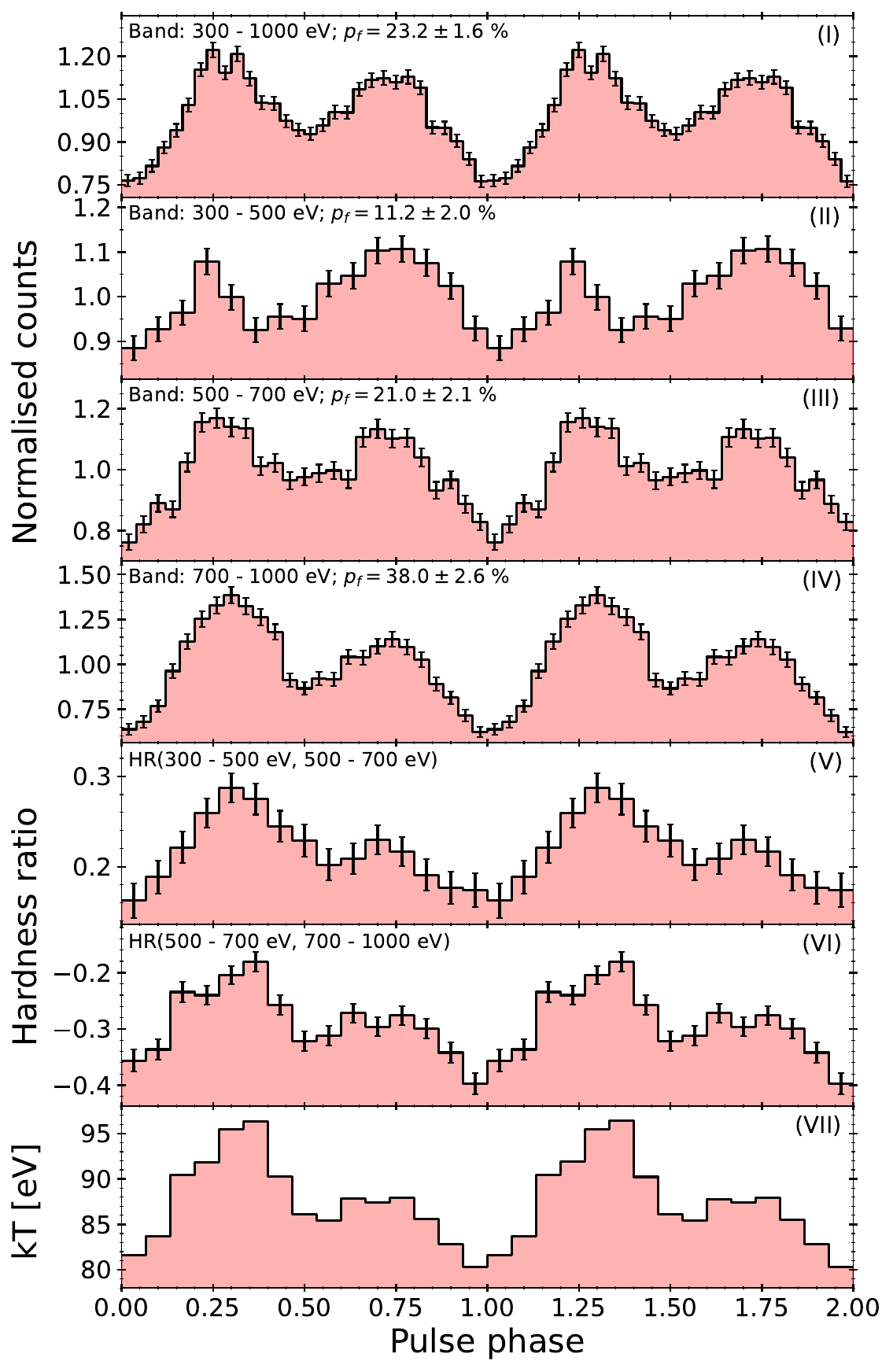}\\
\caption{panel I--IV: Energy-resolved pulse profiles of \otos, combining photons from all observations. panel V and VI: Phase-resolved hardness ratio profiles. panel VII: Phase evolution in mean surface temperature.}
\label{fig_pulse_prof}
\end{figure}


\begin{table*}
\caption{Results of the X-ray spectral modelling.
\label{tab_fitres}}
\centering
\scalebox{.83}{
\begin{tabular}{cccccccccccccc}
\hline\hline
 & $\nh$ & $N_\mathrm{H, gal}$\tablefootmark{(a)} & $kT$ & $R$\tablefootmark{(b)} & $\epsilon_1$\tablefootmark{(c)} & $\sigma_1$ & $EW_1$\tablefootmark{(d)} & $\epsilon_2$\tablefootmark{(c)} & $\sigma_2$ & $EW_2$\tablefootmark{(d)} & $\chi^2_\nu(\nu)$ & Absorbed flux\tablefootmark{(e)}\\
 &  $[10^{20}$\,cm$^{-2} ] $ &  $[10^{20}$\,cm$^{-2} ] $ & [eV] & [km] & [eV] & [eV] & [eV] & [eV] & [eV] & [eV] & & [$10^{-13}$\,\fluxcgs]\\
\hline\noalign{\smallskip}
EPIC & $<6$ & $6.68$ & $87.9^{+1.7}_{-2.9}$ & $7.1^{+2.1}_{-0.8}$ & $270^{+80}_{-5}$ & $147^{+26}_{-40}$ & $420^{+120}_{-120}$ & $596^{+12}_{-11}$ & $47^{+18}_{-16}$ & $45^{+29}_{-17}$ & 1.19(75) & $4.74^{+0.05}_{-0.05}$\\
All & $3.8^{+1.0}_{-0.9}$ & $6.68$ & $86.7^{+1.6}_{-1.9}$ & $7.9^{+1.2}_{-0.8}$ & $<250$ & $208^{+7}_{-18}$ & $450^{+50}_{-170}$ & & & & 1.25(622) & $4.608^{+0.024}_{-0.024}$\\
All & $<0.8$ & $6.68$ & $88.2^{+1.9}_{-2.1}$ & $6.9^{+1.7}_{-0.8}$ & $<240$ & $150^{+120}_{-40}$ & $400^{+180}_{-250}$ & $<590$ & $120^{+60}_{-60}$ & $<190$ & 1.25(619) & $4.635^{+0.024}_{-0.024}$ \\
\noalign{\smallskip}\hline
\end{tabular}}
\tablefoot{Errors correspond to $1\sigma$ confidence intervals.
\tablefoottext{a}{Galactic hydrogen column density in the direction of \otos. Inferred from \citet{2016A&A...594A.116H}.}
\tablefoottext{b}{Blackbody emission radius at infinity, computed assuming a distance of 1\,kpc.}
\tablefoottext{c}{Central energy of the Gaussian absorption component.}
\tablefoottext{d}{Equivalent width (EW), estimated as $\int \frac{f_c-f_o}{f_c} dE$, where $f_c$ denotes the continuum and $f_o$ the observed flux. Errors correspond to the maximum and minimum EW values obtained from all combinations of the upper and lower $1\sigma$ confidence interval limits of the model parameters.}
\tablefoottext{e}{Absorbed model flux in the $0.2-12$\,keV range.}
}
\end{table*}

Compared to \citet{2024A&A...683A.164K}, the additional \nicer\ monitoring, together with the phase-coherent timing solution, enables a more detailed study of the pulse profile. In Fig.~\ref{fig_pulse_prof}, we present pulse profiles normalised by the mean number of counts for varying energy bands, computed from the \nicer\ and EPIC pn observations. To account for the substantial background contamination in the \nicer\ data (Table~\ref{tab_obs}), we applied the best-fitting SCORPEON models (see Sect.~\ref{sec_spect}) to subtract the background. We note, however, that this subtraction may be imperfect at the softest energies, as the SCORPEON models were fitted only above 400~eV and the \nicer\ background increases towards lower energies. For EPIC pn, the expected background contribution was subtracted based on the extracted background spectrum. The MOS data were not included in this analysis, as the significantly lower number of counts per spectral bin would introduce disproportionately large uncertainties in the folded light curves. Finally, the individual data sets were combined by weighing them according to their respective exposure times and relative effective areas.

We confirm the results of \citet{2024A&A...683A.164K} that the pulsed fraction increases with energy from $\sim10$~\% at 0.3--0.5~keV to $\sim40$~\% in the 0.7--1~keV band (Fig.~\ref{fig_pulse_prof}). In addition, no significant variations in the pulsed fractions are observed between the individual observations. The improved signal-to-noise ratio now allows, for the first time, a detailed investigation of the energy dependence of the pulse profile.  Specifically, the first pulse, centred around phase 0.25 (Fig.~\ref{fig_pulse_prof}), increases more strongly in relative amplitude with energy than the second pulse at phase $\sim0.75$. Moreover, the first pulse appears to broaden and its maximum shifts towards later phases at higher energies. In contrast, the second pulse shows no measurable evolution in either phase or width, within the accuracy afforded by the current data.

We computed the hardness ratio (relative difference in detected counts between neighbouring energy bands, calculated via $HR(B1,B2) = \frac{CTS_{B2}-CTS_{B1}}{CTS_{B1}+CTS_{B2}}$) from the pulse profiles over the course of \otos's rotation. The resulting curves are presented in panel V and VI in Fig.~\ref{fig_pulse_prof}. From them, we found \otos's spectrum to harden during the pulses, which is strongest for the pulse around phase $0.25$. Assuming that the hardening of \otos's spectrum is solely caused by a change in mean surface temperature, for each phase bin in Fig.~\ref{fig_pulse_prof} we applied the best-fit model to the EPIC spectra (see Sect.~\ref{sec_spect} and first row in Table~\ref{tab_fitres}) and the XSPEC \citep[version: 12.15.0;][]{1996ASPC..101...17A} tool to simulate spectra with varying blackbody temperatures to find the temperature that best reproduces the observed hardness ratio values. The resulting evolution of the best-fit temperature values over \otos's rotation is presented in panel VII in Fig.~\ref{fig_pulse_prof} and indicates the mean surface temperature to vary between $\sim80$~eV and $\sim 96$~eV.

\subsection{Spectral analysis \label{sec_spect}}

\begin{figure}[t]
\includegraphics[width=1.02\linewidth]{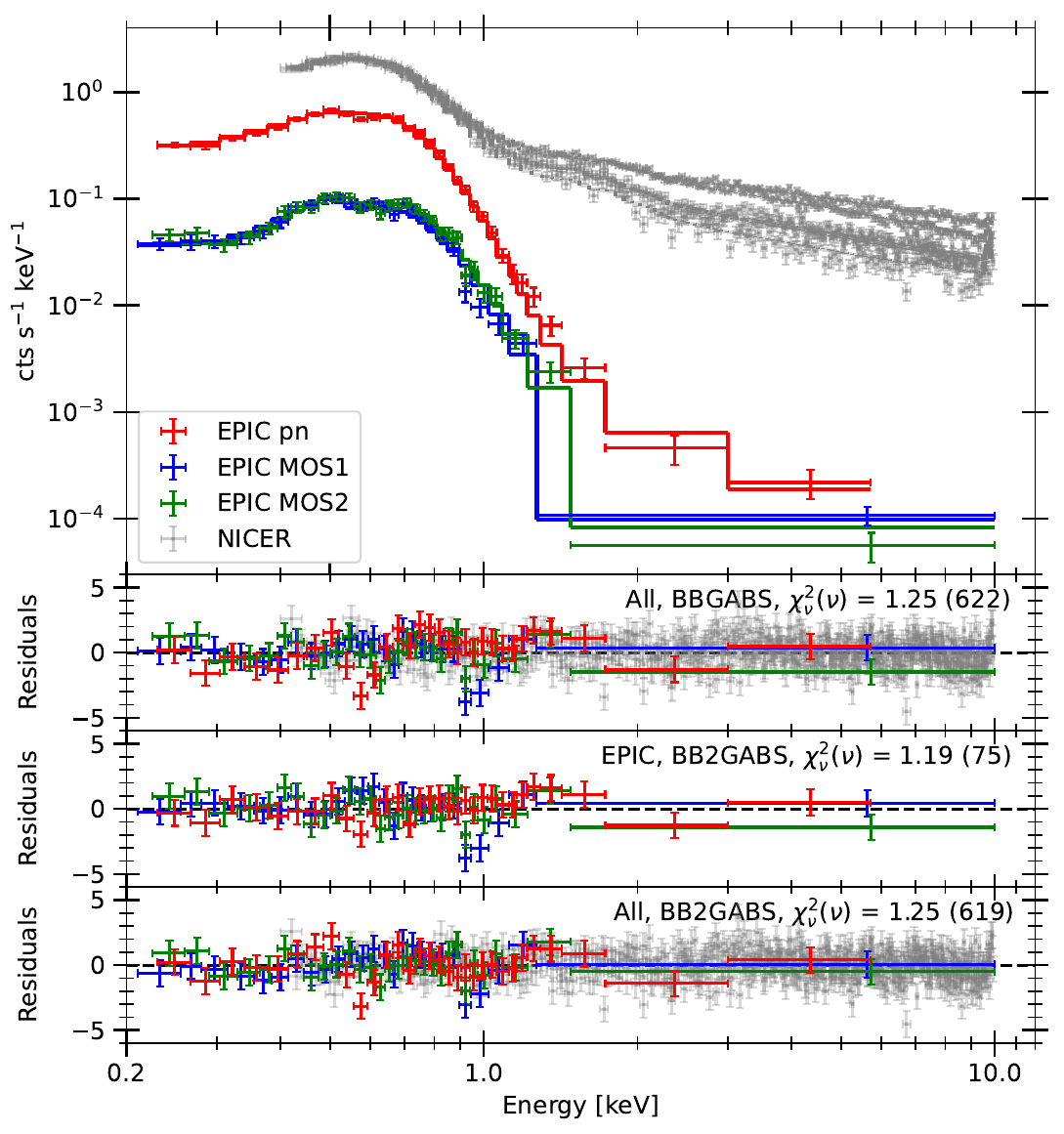}\\
\caption{\xmm\ and \nicer\ X-ray spectra along with the three models presented in Table~\ref{tab_fitres}. The upper panel presents the best-fit single-blackbody and Gaussian absorption line model (\textit{\mbox{BBGABS}}) that was simultaneously fitted to all the spectra (second line in Table~\ref{tab_fitres}).}
\label{fig_specplot}
\end{figure}

We applied the X-ray spectral fitting tool XSPEC to study the phase-averaged spectral emission of \otos. For all fits, we adopted the $\chi^2$ statistic and included \texttt{tbabs} components to account for interstellar absorption, using elemental abundances from \citet{2000ApJ...542..914W}. Following the approach of \citet{2024A&A...683A.164K}, we consistently added an absorbed power law component to the spectral model to account for contamination from a nearby hard X-ray emitter to the north of \otos. For this component, we fixed the hydrogen column density to $\nh = 10^{21}$~cm$^{-2}$ and the photon index to $\Gamma = 1.81$, leaving only the normalisation free to vary between the individual spectra, to allow for different levels of contamination. We further included a multiplicative constant to account for residual cross-calibration uncertainties between the instruments. The background in the individual \nicer\ spectra was modelled by simultaneously fitting \textit{SCORPEON}\footnote{\url{https://heasarc.gsfc.nasa.gov/docs/nicer/analysis_threads/scorpeon-overview/}} models. The fit parameters for the best-fit models are listed in Table~\ref{tab_fitres}, whereas the spectra and residuals are presented in Fig.~\ref{fig_specplot}.

The results of the spectral modelling are consistent with those reported in \citet{2024A&A...683A.164K}. Fitting only the EPIC spectra (EPIC pn, MOS1, MOS2), we found that a single absorbed blackbody of $kT\sim 88$~eV, superimposed by two absorption features at $\sim 270$~eV and $\sim595$~eV, provides the best fit (Table~\ref{tab_fitres}).

Fits to the individual \nicer\ spectra indicate that consistent results with \xmm\ can only be achieved if the spectral range is restricted at the soft end; accordingly, we used a spectral range of $0.4-10$~keV, instead of the full $0.25-10$~keV. The spectral parameters obtained from the individual \nicer\ fits agree within $1-2\sigma$, indicating the absence of significant spectral state changes in \otos\ over the course of the observational campaign.

When fitting all spectra (\xmm\ and \nicer), we found that the inclusion of the \nicer\ data makes it difficult to recover the absorption line at $\sim595$~eV. As summarised in Table~\ref{tab_fitres}, adding a second absorption feature does not improve the fit statistic beyond that obtained with a single line, and the parameters of the second feature are poorly constrained. We attribute this to the larger noise in the \nicer\ spectra, that dominate the spectral fit at the energy of the second feature relative to the EPIC data.

The inclusion of further model components, such as additional blackbody components to model hot spots or a power law to account for possible high-energy emission from \otos, does not improve the fits beyond those reported in Table~\ref{tab_fitres}. For instance, the light curve analysis in Sect.~\ref{sec_timing} implies the surface temperature to vary between 80~eV and 96~eV, if modelled with a single blackbody continuum. We attempted to investigate the surface temperature variations in more detail by fitting a model consisting of two blackbody components and one absorption feature, were we fixed the blackbody temperatures to 80~eV and 96~eV, respectively. While a simultaneous fit of all the \nicer\ and EPIC spectra with this model is of similar quality ($\chi^2_\nu (\nu) = 1.25(622)$) as the single blackbody continuum fit presented in Table~\ref{tab_fitres}, this model converges to a large $\nh=1.3^{+0.5}_{-0.4}\times 10^{21}$~cm$^{-2}$, in excess of the Galactic value, and comparably worse fit statistic ($\chi^2_\nu (\nu) = 1.32(78)$), if it is applied only to the EPIC data. Leaving the blackbody temperatures as free parameters in the EPIC fit allows to improve the fit statistic ($\chi^2_\nu (\nu) = 1.09(76)$), but also converges to unreasonably large $\nh = 1.4^{+0.5}_{-0.4}\times 10^{21}$~cm$^{-2}$ and radius ($R>160$~km) values. Such a result was already reported for fits to EPIC pn alone in \citet{2024A&A...683A.164K}. We envisage to conduct and present more detailed spectroscopic studies of \otos\ in a future publication.

Comparing the hydrogen column density with Galactic absorption maps provides one of the few methods to estimate distances for predominantly thermally emitting INSs. As the \nicer\ spectra are restricted at soft energies, the degeneracy between interstellar absorption and the low-energy absorption feature cannot be fully resolved, making it difficult to constrain $\nh$ precisely. Using the 1$\sigma$ upper limits on $\nh$ and the 3D-$\nh$ tool \citep{2024arXiv240303127D}, we derived maximum distances of 650--700~pc for both the two-line fit to the EPIC data and the single-line fit to all spectra. Assuming a lower distance limit of 100~pc, this corresponds to a thermal luminosity in the range $3\times 10^{30}$~erg~s$^{-1}$ to $3\times 10^{32}$~erg~s$^{-1}$. These values are consistent with the thermal luminosities observed in middle-aged INS classes such as XDINSs or RPPs \citep[e.g.][]{2020MNRAS.496.5052P}.

\section{Discussion\label{sec_disc}}

The \nicer\ and \xmm\ observations of \otos, spanning nearly 15 months, provide further insight into the nature of this neutron star. These data not only allow a more precise measurement of the spin period, but also enable, for the first time, a determination of the source's spin-down rate, $\dot{P}=8.8^{+1.7}_{-1.6} \times 10^{-14}$~s~s$^{-1}$ (Table~\ref{tab_tim}). This result is important, as the timing properties of INSs relate to their physical properties and evolution. The timing solution obtained in this work allows to coherently connect all X-ray observations and to accurately keep track of \otos's cycle counting.

\begin{figure}[t]
\includegraphics[width=1.02\linewidth]{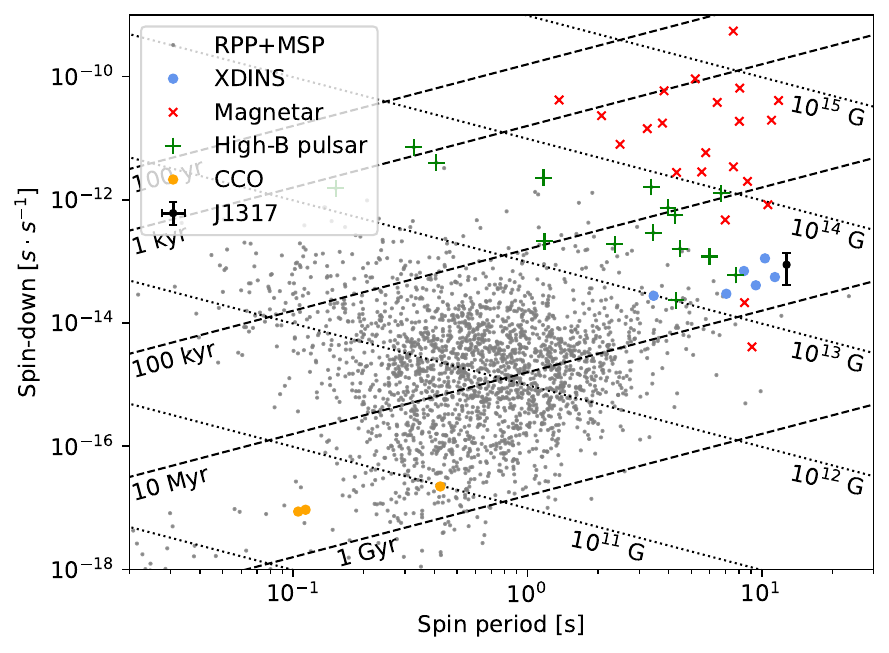}\\
\caption{Spin period versus spin-down diagram showing the location of \otos\ relative to INSs listed in the ATNF pulsar catalogue \citep[][]{2005AJ....129.1993M}.}
\label{fig_ppdot}
\end{figure}

Locating \otos\ in a spin-period versus spin-down diagram (Fig.~\ref{fig_ppdot}) alongside all INSs listed in the ATNF pulsar catalogue \citep[version: 2.7.0;][]{2005AJ....129.1993M} helps to constrain the INS type. Situated at the high-period end of the INS population, \otos\ occupies a region predominantly populated by XDINSs, high magnetic field RPPs, and few magnetars. The dipolar magnetic field strength inferred from the timing solution under the assumption of pure dipolar magnetic braking, $B= 3.4^{+0.4}_{-0.4} \times 10^{13}$~G, is typical for XDINSs or high-B RPPs, whereas the timing properties of most magnetars imply even stronger dipolar fields \citep[e.g.][]{2017ARA&A..55..261K}.

\otos's dipolar magnetic field strength is consistent within two sigma with the magnetic field inferred from the broad absorption feature in its spectrum \citep[e.g. $5.6^{+1.8}_{-1.1}\times 10^{13}$~G from a fit to EPIC pn;][]{2024A&A...683A.164K}, assuming the feature arises from proton cyclotron resonances. This is reminiscing of the XDINS \magot, whose timing properties imply a similarly strong dipolar magnetic field of $\sim 3.4\times 10^{13}$~G \citep{2024ApJ...969...53B} and which possesses a broad absorption feature at $203^{+40}_{-37}$~eV \citep{2003A&A...403L..19H,2017MNRAS.468.2975B} that, assuming proton cyclotron resonances to be the cause, implies a consistent magnetic field strength of $4.3^{+0.9}_{-0.8}\times 10^{13}$~G. It is, however, important to note that such an agreement in field strength is not a general occurrence for XDINSs. For the other five XDINSs with detected pulsations, the timing properties imply lower dipolar magnetic fields in the range of $(1-2)\times10^{13}$~G \citep{2024ApJ...969...53B}, whereas the absorption features are typically detected at energies $\gtrsim300$~eV, implying stronger fields. A possible way to resolve this conundrum could be, if the features originate from smaller high-magnetic field structures, which could also account for the phase variable nature observed in some of the lines \citep[e.g.][]{2017MNRAS.468.2975B}. Alternatively, apparent absorption features may also be introduced from a non-perfect spectral modelling of a more complex surface temperature distribution \citep{2014MNRAS.443...31V,2024ApJ...969...53B}.

The characteristic age of \otos, $\tau=2.3^{+0.5}_{-0.4}$~Myr, is similarly typical for XDINSs or high-B RPPs. It should be noted, however, that the characteristic age may strongly deviate from the true age, as for example observed in the XDINSs \citep[e.g.][]{2010MNRAS.402.2369T}, if the spin-evolution is not solely governed by pure dipolar magnetic braking. Additional processes, such as magnetic field decay, the presence of a fallback disc, particle outflows, or gravitational wave emission, may influence angular momentum loss \citep[e.g.][]{1969Natur.223..277M,1996A&A...312..675B,1998PhRvD..58h4020O,2013MNRAS.434..123V,2014MNRAS.444.1559E,2024MNRAS.534.1481G}. In principle, a measurement of \otos's braking index would allow one to probe whether any of these mechanisms are at work; however, this requires determination of the second period or frequency derivative, which the current data do not yet permit.

\begin{figure}[t]
\includegraphics[width=1.02\linewidth]{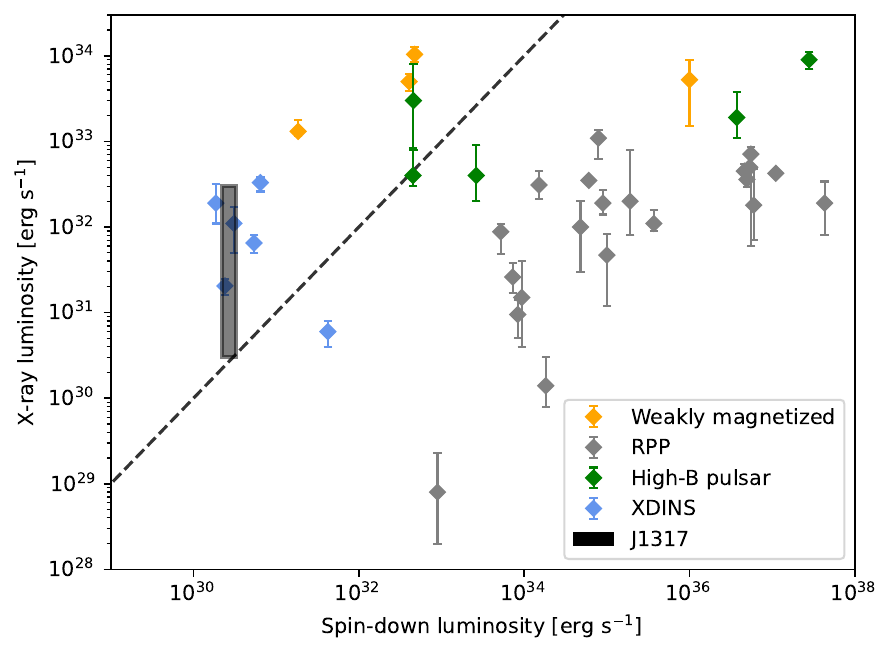}\\
\caption{Thermal luminosity as a function of spin-down luminosity for different INSs \citep{2020MNRAS.496.5052P}. The location of \otos\ is marked by the black box. The spin-down luminosity is derived from the timing properties (see notes of Table~\ref{tab_tim}) of \otos, while the upper and lower thermal luminosity limits correspond to an assumed distance range of 100--700~pc.}
\label{fig_lum}
\end{figure}


Rotation-powered pulsars typically exhibit spin-down luminosities that exceed their thermal luminosities, whereas predominantly thermally or magnetically powered INS classes, such as XDINSs, central compact objects in supernova remnants \citep[CCOs;][]{2017JPhCS.932a2006D}, and magnetars, show thermal luminosities that often surpass their rotational energy losses \citep[e.g.][]{2019RPPh...82j6901E, 2020MNRAS.496.5052P}. The comparatively strong thermal emission observed in these sources is commonly attributed to internal or external reheating processes, for example driven by magnetic field decay or accretion from a fallback disc \citep{2010A&A...522A..16G,2013MNRAS.434..123V,2014MNRAS.444.1559E}. Thus, a comparison between thermal luminosity and spin-down luminosity provides a valuable probe of \otos's past thermal evolution and INS nature. In Fig.~\ref{fig_lum}, we show the thermal and spin-down luminosities of different INS classes compiled by \citet{2020MNRAS.496.5052P}. Although the distance to \otos\ remains uncertain, owing to the difficulties in modelling the soft part of its X-ray spectrum, even at a distance as small as 100~pc, \otos's thermal luminosity is comparable to its rotational energy loss. This places the source in a region of the diagram predominantly occupied by XDINSs, supporting such an interpretation. The relatively strong thermal luminosity of \otos\ further suggests that reheating processes have influenced its past thermal evolution. At larger distances ($>1$~kpc), exceeding the $1\sigma$ upper distance limit inferred in this work, the thermal luminosity could become consistent with that of magnetars or CCOs. We note, however, that the best-fitting blackbody temperature of \otos\ ($\sim90$~eV) would be unusually low for these younger INS classes.

While XDINSs are generally stable X-ray emitters, long-term monitoring of the original seven \ros-discovered sources has revealed spectral or timing variability for some of them. The most notable example is \magzs, which underwent a glitch event accompanied by a change in its X-ray spectrum \citep{2012MNRAS.423.1194H, 2024ApJ...969...53B}. More recently, variations in spectral state were reported for \magos\ \citep{2023IAUS..363..288M}, while a change in timing state reported for \magto\ could not be confirmed \citep{2026ApJ...996...79P}. No significant changes in spectral flux or best-fitting parameters are observed in the individual fits to the \nicer\ spectra of \otos, indicating overall stable X-ray emission. Likewise, the fact that a simple quadratic timing solution adequately describes the timing evolution of \otos, together with the absence of measurable changes in pulsed fraction between individual epochs, suggests that no major timing irregularities, such as glitches, occurred during the observational baseline. This result may be expected, if one exemplarily considers that only two changes in spectral or timing state have been observed in the \ros\ discovered XDINSs, although their whole observational baseline (adding up the time between the first and most recent \chan, \nicer, or \xmm\ observation of each XDINS) comprises almost 60000 days. The $\sim450$~d between the start and end of the first and last \nicer\ observations of \otos\ are very short in comparison.

It was already noted in \citet{2024A&A...683A.164K} that \otos's spectral and timing properties are strikingly similar to those of the XDINS \magot. In both sources, a broad absorption feature is required to model the spectrum at the soft end and both exhibit the strongest pulsations among XDINSs (e.g. we measured \otos's pulsed fraction to be $23.2 \pm 1.6 $\% in the 0.3--1~keV band). Recently, \citet{2024ApJ...969...53B} studied the pulse profiles of the six \ros\ discovered XDINSs with detected modulations, noting that \magot\ has a double-humped profile, with one pulse broader than the other. While both pulses in \magot\ have comparable amplitudes over the soft X-ray band, their relative strengths change with energy: the broad pulse grows and shifts about 0.05 rotational cycles forward at higher energies, whereas the narrow pulse is slightly stronger at lower energies and diminishes above 0.8~keV. The pulse-behaviour observed in \otos\ is similar (Fig.~\ref{fig_pulse_prof}). At 0.3--0.5~keV, the pulse immediately following the lowest minimum around rotational phase 0.25 appears narrower than the second pulse around phase 0.75. Moving to higher energies, the second pulse does not significantly change, while the formerly narrow first pulse becomes more dominant, increasing in amplitude and width. This produces an overall increase in pulsed fraction with energy (from $\sim10$~\% at 0.3--0.5~keV to $\sim40$~\% in the 0.7--1~keV band; Fig.~\ref{fig_pulse_prof}). Similar to \magot, the maximum of \otos's first pulse appears to shift, although in the opposite direction by $\sim0.05$ rotational cycles to later phases.

Pulsations in XDINSs are generally thought to arise from a combination of factors: a complex hot-spot distribution, non-isotropic radiation as might be produced by an atmospheric layer, or energy- and phase-dependent absorption processes, such as cyclotron resonances \citep{2024ApJ...969...53B}. The similarities between \otos\ and \magot\ therefore suggest that both sources may share overall comparable surface structures and/or emission properties, as well as a similar line of sight. In \citet{2024A&A...683A.164K} the EPIC pn observation was used to study \otos's phase-resolved spectral properties. While this analysis showed the broad absorption feature at $\sim300$~eV to be observable at all phases, the single EPIC pn observation alone does not permit to constrain the phase dependence of the narrow feature at $\sim590$~eV, nor to study \otos's rotational temperature evolution in more detail. It is, however, clear that the absorption features alone can not cause the observed modulations in \otos, as the pulsations grow in strength towards higher energies, where no spectral features are observed.

Reconstructing \otos's surface temperature profile, to directly investigate whether a complex hot-spot distribution causes the pulsations, is even with the new \nicer\ data still difficult to do, as high signal-to-noise light curves are required for this task. Nevertheless, the rotational hardness ratio evolution (panels V and VI in Fig.~\ref{fig_pulse_prof}) can be used to obtain estimates on the change in mean surface temperature over the course of the neutron star's rotation, assuming that \otos's modulations are solely caused by a non-uniform surface temperature distribution. The rotational evolution of mean surface temperature is presented in Fig.~\ref{fig_pulse_prof}, indicating it to vary between $\sim80$~eV and $\sim 96$~eV. While these values are within the typical temperature range observed in XDINSs, even if additional spectral components for heated surface regions are accounted for \citep[e.g. see Table~3 and 5 in][]{2019PASJ...71...17Y}, higher signal-to-noise phase-resolved studies of \otos\ are still warranted, as the temperature variation of a single blackbody is a very simple approximation.

That \otos\ possesses a complex surface temperature distribution may, nevertheless, be already implied from its strong pulsation. Assuming isotropic emission and simulating the magneto-thermal evolution of INSs with different magnetic field configurations, \citet{2013MNRAS.434.2362P} found that dipolar magnetic field configurations only allow for symmetric pulse profiles and low pulsed fractions. Conversely, if the magnetic field has strong toroidal components, anisotropic surface temperature profiles and large pulsed fractions can be observed, similar to those observed for \otos. We note that the phase shift of the pulse maximum with energy could also hint towards the presence of toroidal field components, as such a phase shift may imply the hot spot(s) to not be located antipodal to each other, as would be expected for a field configuration that is dominated by dipolar components. A non-antipodal location of hot spots was also proposed for \magot\ based on fitting energy resolved light curves \citep{2005A&A...441..597S}.

Alternatively, magnetised INS atmospheres can also cause anisotropic emission that leads to pulsations and pulse profiles reminiscent of those observed in the XDINSs \citep[e.g.][]{2006MNRAS.366..727Z,2013MNRAS.434.2362P}. \citet{2011A&A...534A..74H} found that \magot's phase-resolved spectra can be well-fitted by modelling a condensed iron surface with a partially ionised hydrogen layer above it. This fit implies the absence of strong toroidal magnetic field components in the source and that \magot\ is an orthogonal rotator. Conducting such studies for \otos\ goes beyond the scope of this work, but could be very interesting, as constraining the viewing geometry of the source would give further insights into the mechanisms that cause the radio-quiet nature of XDINSs. It is not clear, if there are intrinsic reasons for the absence of strong magnetospheric emission in these sources or if the radio-quiet nature is caused by an unfavourable viewing geometry, where the narrow radio beam misses the observer \citep[e.g.][]{2009ApJ...702..692K}. That geometric reasons may be at fault is for instance indicated from the high-B pulsar \zsts. While \zsts's X-ray emission properties are overall similar to \magot\ and \otos\ (predominantly thermal X-ray spectrum, broad absorption feature at soft X-ray energies, double-peaked pulse profile, pulsed fraction increasing with energy, etc.), the detection of radio pulsations in the source and results from the detailed phase-resolved analysis conducted in \citet{2019A&A...627A..69R} indicate a lower impact parameter (absolute deviation of the angles between the rotation axis and line-of-sight or dipole axis) in respect to the known XDINSs \magzs\ and \magot. If these results are taken at face value, they imply that the narrow radio beam simply misses the observer in the XDINS case, whereas the geometry of \zsts\ allows for its detection. Constraining the impact parameter for \otos\ would allow to further probe whether geometric reasons cause the absence of detectable radio emission in the XDINSs or if there are other physical mechanisms at work. Likewise, deep radio follow-up observations of \otos\ could be of interest to search for detectable radio emission in the source.

The results of \otos's X-ray spectral modelling are overall consistent with \citet{2024A&A...683A.164K}, although higher signal-to-noise spectra, particularly below 0.6~keV, would be helpful to lift remaining degeneracies and better constrain the line parameters. The fact that a simple absorbed blackbody fits \otos's continuum well is consistent with an XDINS nature, as these sources are characterised by predominantly thermal blackbody-like X-ray emission. While non-thermal emission components have been identified in two XDINSs \citep[\magzf, \magoe;][]{2020ApJ...904...42D,2022MNRAS.516.4932D}, high energy emission remains elusive in the other sources.

To better quantify the upper limit on hard X-ray emission in \otos, we applied the point-spread-function (PSF) fitting technique described in Sect.~3.2 of \citet{2026A&A...705A.148K} to the EPIC pn observation. Count rates were converted to flux using the energy conversion factor (ECF) computed from the extracted EPIC pn response files, assuming a power law high-energy spectrum with $\nh=6\times 10^{20}$~cm$^{-2}$ and $\Gamma=2$. The PSF fit at \otos's position was performed simultaneously with PSF fits to nearby sources detected in a previous source detection run in the 1.5--12~keV energy band. We determined a flux of $5.2^{+2.4}_{-2.4}\times 10^{-15}$~\fluxcgs\ (1.5--12~keV) at \otos's position. Given that this flux is consistent with zero within 2--3$\sigma$ and the source is not formally detected ($DET\_ML=2.1$), we conclude that no high-energy emission from \otos\ is detected in EPIC pn. This implies an upper limit on the non-thermal to thermal flux ratio of $1.7^{+0.9}_{-0.9}$\%, which remains shallow compared to the ratios observed in the two XDINSs with detected hard X-ray excess \citep[e.g. see Sect~3.2 in][]{2026A&A...705A.148K}. Deeper X-ray observations are required to confirm or rule-out magnetospheric emission from the source.

\section{Summary and conclusions\label{sec_concl}}

The new \nicer\ epochs, in combination with previous X-ray follow-up, allowed us to further investigate \otos's INS nature. Based on its timing properties and relatively high thermal luminosity, we classify the source as a member of the XDINS class. This conclusion is further supported by the inferred dipolar magnetic field strength of $\sim3\times 10^{13}$~G, its predominantly thermal spectrum, the absence of detectable high-energy emission, and pulse properties that closely resemble those of the XDINS \magot. The characterisation of the source can still benefit from additional X-ray observations, for instance to extend the validity of the newly established coherent timing solution. Higher signal-to-noise spectra are also required to lift the degeneracy between the broad absorption feature and interstellar absorption, which is necessary to better understand the line-forming mechanism and obtain more accurate distance estimates. Nonetheless, the confirmation of \otos's XDINSs nature marks the first XDINS discovery in more than two decades and represents an important step towards a larger sample of these enigmatic neutron stars.

\begin{acknowledgements}
We thank the anonymous referee for their helpful feedback and comments that allowed us to furher improve the quality of this work.

This work was funded by the Deutsche Forschungsgemeinschaft (DFG, German Research Foundation) through grants Schw 536/38-1 and Schw 536/38-2, and by Deutsches Zentrum für Luft- und Raumfahrt (DLR) through grant Fkz 50\,OR\,2408.

AMP acknowledges the Innovation and Development Fund of Science and Technology of the Institute of Geochemistry, Chinese Academy of Sciences, the National Key Research and Development Program of China (Grant No. 2022YFF0503100), the Strategic Priority Research Program of the Chinese Academy of Sciences (Grant No. XDB 41000000), and the Key Research Program of the Chinese Academy of Sciences (Grant NO. KGFZD-145-23-15).

This research has made use of data and/or software provided by the High Energy Astrophysics Science Archive Research Center (HEASARC), which is a service of the Astrophysics Science Division at NASA/GSFC.

This work made use of Astropy (\url{http://www.astropy.org}): a community-developed core Python package and an ecosystem of tools and resources for astronomy \citep{astropy:2013, astropy:2018, astropy:2022}.

We derive posterior probability distributions and the Bayesian evidence with the nested sampling Monte Carlo algorithm MLFriends \citep{2014A&A...564A.125B,2019PASP..131j8005B} using the \mbox{UltraNest} (\url{https://johannesbuchner.github.io/UltraNest/}) package \citep{2021JOSS....6.3001B}.
\end{acknowledgements}
\bibliographystyle{aa}
\bibliography{ref_list}
\begin{appendix}
\section{Times-of-arrival measurements\label{sec_appa}}

To conduct the timing searches and make best-use of the \nicer\ observing pattern, consisting of multiple individual short exposures that at each epoch form a full observation, the data from the five \nicer\ observations was split into individual segments containing at least 4000 counts each. In Table~\ref{tab_toas}, we present these individual segments, the number of counts, and respective TOA measurements based on the best-fit quadratic timing solution described in Sect.~\ref{sec_timing}.

\begin{table}[ht!]
\small
\caption{Individual observation segments and TOA values.\label{tab_toas}}
\centering
\scalebox{1.}{
\begin{tabular}{lrrlrrcc}
\hline\hline\noalign{\smallskip}
Instrument & Events & Revolutions & TOA\\
\hline
NICER & 4132 & -2438334 & 60032.823481(4)\\
NICER & 5053 & -2435003 & 60033.315312(4)\\
NICER & 4211 & -2431107 & 60033.890563(6)\\
NICER & 4177 & -2427798 & 60034.379142(6)\\
NICER & 4002 & -2426256 & 60034.606823(6)\\
NICER & 4844 & -2423033 & 60035.082704(5)\\
EPIC pn & 1819 & -1754544 & 60133.7864406(29)\\
EPIC MOS 1 & 1960 & -1754542 & 60133.786735(4)\\
EPIC MOS 2 & 9254 & -1754459 & 60133.7989913(15)\\
NICER & 4098 & 4562 & 60393.521986(5)\\
NICER & 4189 & 10897 & 60394.457359(4)\\
NICER & 4441 & 14822 & 60395.036895(4)\\
NICER & 9565 & 46895 & 60399.7725387(16)\\
NICER & 6305 & 259916 & 60431.2255219(19)\\
NICER & 7147 & 624342 & 60485.0337591(20)\\
\hline
\end{tabular}}
\end{table}

\end{appendix}
\end{document}